\documentclass[pre,amsmath,aps,twocolumn,superscriptaddress,letterpaper]{revtex4}
\usepackage{graphicx,xspace,units,subfigure}
\usepackage{float}
\usepackage{amsmath}
\usepackage{amssymb}
\usepackage[normalem]{ulem}
\input{epsf}

\newcommand{\noi}{\noindent}
\newcommand{\be}{\begin{equation}}
\newcommand{\ee}{\end{equation}}

\begin{document}

\title{Langevin formulation for  single-file diffusion}
\author{Alessandro Taloni}
\email{taloni@phys.sinica.edu.tw}
\affiliation{
Massachusetts Institute of Technology, Department of
  Physics, 77 Massachusetts Avenue, Cambridge, MA 02139, USA}
\affiliation{
   Institute of Physics, Academia Sinica, 128 Academia Road, Taipei
  11529, Taiwan }
\author{Michael A. Lomholt}
\affiliation{
  MEMPHYS - Center for Biomembrane Physics, Department of Physics and Chemistry, 
University of Southern Denmark, Campusvej 55, 5230 Odense M, Denmark}
\affiliation{
Physics Department, Technical University of Munich, D-85747 Garching, Germany}

\begin{abstract}   

We introduce a stochastic equation for the microscopic motion of a
tagged particle in the single file model. This equation provides a
compact representation of several of the system's properties such as
Fluctuation-Dissipation  and Linear Response relations, achieved
by means of a diffusion noise approach. Most important, the proposed Langevin
Equation reproduces quantitatively the \emph{three} temporal regimes and
the corresponding time scales: ballistic, diffusive and
subdiffusive.

\end{abstract}
\maketitle

\section{Introduction}
\label{sec:Introduction}

Since its  introduction in 1965 due to  Harris' pioneering work
\cite{Harris}, the single file model (SF) has attracted more and
more  interest among the scientific community. Introduced 
first in the mathematical physics literature as an interesting though
somewhat exotic topic, it has inspired over the last
40 years a large body of profound theoretical studies
\cite{SF_basic} and detailed 
numerical investigations, including extensive Monte-Carlo
and molecular dynamics simulations \cite{SF_numerics}.

The
motivations for the longstanding interest in this topic  reside, on one
hand, in its analytical tractability and, on the other hand, in its
effectiveness as a description of diffusion phenomena in real
\emph{quasi} one dimensional systems. As a matter of fact, since the
direct observation 
and manipulation of nanoscopic systems has exponentially evolved
in the last decade, models suitable to account for the single particle
diffusional mechanisms
in constrained flow geometries,  have been subjects of increasing
attention.  Remarkably, among these,  the SF  holds a preeminent
position, since it correctly reproduces  
transport properties in a large category of \emph{quasi} one
dimensional  systems, where each particle is free to diffuse against
its neighbors but is forbidden to overcome them \cite{SF_modelz}. 
Transport processes of this type may be observed  in nanoporous
materials \cite{Karger_Book}, in collective 
motion of ions through biological  channels and membranes
\cite{Aidley_Book} as well as  in nanodevices and cellular flows
\cite{Albert_Book}

Along a mathematical line, the SF is perhaps the 
simplest interacting  one dimensional gas one can consider: it consists of N
unit-mass particles constrained to move along  a line following
a given 
dynamics. As the particles' interaction is purely hard-core, no mutual
exchanges of the diffusants are allowed, i.e. they retain
their ordering over time (\emph{single filing condition}). In
spite of the intricacies of its mathematical derivation, the long
time behavior of the single
file dispersion relation can be cast in  
the following suggestive form  \cite{SF_basic}

\be
\delta x^2(t)=\frac{\langle \left|X\right|\rangle}{\rho}
\label{SFdisp}
\ee

\noi where $\rho=\frac{N}{L}$ ($N\to\infty$, $L\to\infty$) is the
file's density and $\langle \left|X\right|\rangle$ is the absolute
displacement of a 
non-interacting particle. If the free particle dynamics is characterized
by a diffusivity $D=k_BT/\gamma$, then the relation (\ref{SFdisp})
takes the form \cite{SF_basic2}

\be
\delta x^2(t)=2\sqrt{\frac{ D\,t}{\pi\rho^2}}.
\label{SFstoc}
\ee

\noi The predicted subdiffusive behavior has been reproduced
experimentally in colloidal particles systems \cite{colloidal} and
 observed in molecular sieves (zeolites)
\cite{zeolites}. 
We remark that the subdiffusive behavior represented by
Eq. (\ref{SFstoc}) will eventually be replaced by regular diffusion if
for instance the particles are allowed to overtake each other
\cite{mon}, if there is only a finite number of particles
\cite{kumasl}, or the particles move in a ring \cite{beijeren}.

In Ref.\cite{Marchesoni} it was pointed out that the subdiffusive regime
of a SF
tagged particle occurs on the score of long ranged
anticorrelations of its velocity, and/or of the jump's statistics of the
collisional mechanism underlying its dynamics. Beside these persistent
memory effects, different mathematical derivations 
 agree with the fact that asymptotically the tagged particle's
probability distribution must be Gaussian with a variance growing in
time according to (\ref{SFstoc}) \cite{SF_basic}. Together, these
properties contrast 
with the Continuous Time Random Walk (CTRW) scheme and its corresponding
Fokker Planck representation, for which a stretched Gaussian solution
has to be expected (see Ref.\cite{Klafter} and references therein).
Furthermore, several  subdiffusive systems in nature share
the property of Gaussianicity  with the SF, e.g. a  monomer in a one dimensional
phantom polymer  
\cite{phantom_monomer}, the ``translocation coordinate'' of a two
dimensional Rouse chain through a hole \cite{Kantor},  a
tagged monomer in an Edward-Wilkinson chain \cite{Alberto}, de
Gennes' defects along a polymer during its reptation \cite{DeGennes}
and solitons in the sine-Gordon chain \cite{solitons}.

 In this paper we address the question of the 
microscopic effective description of the stochastic, anomalous motion
of the tagged particle. Our aim is to extend the valuable Langevin
approach, valid for the case of a diffusive Brownian walker, to the
subdiffusive dynamics of a SF particle. We anticipate here that the
Generalized Langevin Equation (GLE) \cite{Mori,Kubo} provides the ideal
theoretical tool for such a goal, incorporating all the statistical
properties enjoyed by the particle. Within this framework the non-Markovian
memory effects are 
achieved by means of a power-law damping kernel \cite{Lutz}, which is
simply added algebraically 
to the instantaneous friction of the surroundings. We note here that  the GLE
has  recently been 
successfully used to describe several physical phenomena and market
flows \cite{GLE_articles}.

The article is organized as follows: in sec.\ref{sec:Diffusion_noise}
we study the density profile dynamics by means of a Diffusion-noise
approach and we connect  the file's density  fluctuations 
to the motion of a tagged particle.  In
sec.\ref{sec:Generalized_Langevin} we introduce the GLE and we show
the accuracy of the Langevin description
by means of extensive molecular dynamic simulations.

\section{Diffusion noise approach}
\label{sec:Diffusion_noise}

We start by considering the file density dynamics. As stated in
\ref{sec:Introduction}, the system is composed
of $N$ Brownian point-like particles, all of unit  mass, moving
along a ring of length $L$ and performing a stochastic motion according to the
Langevin Equation (LE):

\be
\begin{array}{l}
\dot{x_i}(t)  =  v_i(t)\\
\dot{v_i}(t)  =  -\gamma v_i(t)+\xi_i(t),
\label{LE}
\end{array}
\ee

\noi
where $i\in[1,N]$ denotes the particle's index, and the damping
coefficient $\gamma$ and the random noise source $\xi_i(t)$ satisfy the
well-known \emph{Fluctuation-Dissipation Relation} (often called second
Kubo Theorem \cite{Kubo})
$\langle\xi_i(t)\rangle\xi_j(t')\rangle=2k_BT\gamma\delta_{i,j}\delta(t-t')$.
 The single filing condition  turns out to be merely the interchange of
two particles labels, whenever these suffer an (elastic)
collision. However, as pointed out in \cite{Jepsen_Lebowitz}, all
system properties which do not depend on particle labeling remain
unchanged from those of an ideal gas, i.e. of N independent Brownian
walkers.

\begin{figure}[t]
\includegraphics[width=6cm]{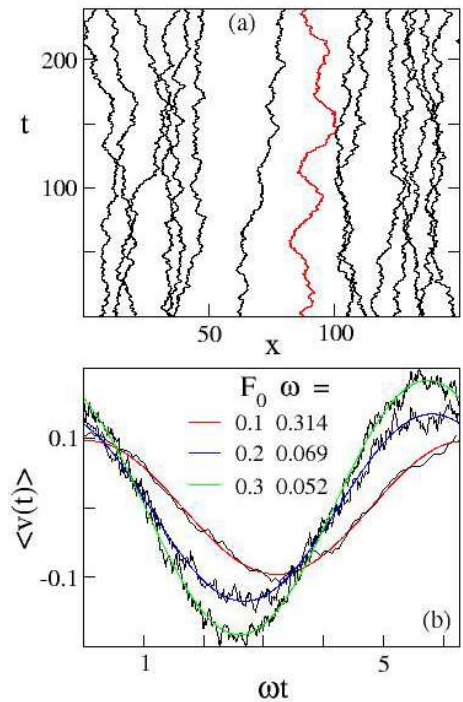}
\caption{(Color online). Panel (a): typical SF dynamics. The tagged
  particle (red) is subjected to the external periodic force
  $F(t)=F_0\cos\omega t$, while the other ones perform the usual
  stochastic dynamics. Panel (b): Linear Response relation
  (\ref{LinearResponse}). Average velocity $\langle v(t)\rangle$
  \emph{vs} $\omega t$ for $k_BT=1.0$, $\gamma=0.5$,
  $\rho=0.25$ and different values of $F_0$ and $\omega$. Three
  typical numerical curves (black lines) are fitted through the
  relation (\ref{LinearResponse}), showing a linear dependence on the
  amplitude of the applied sinusoidal force. Averages have been taken
  over more than 2000 periods of each realization, performing at least 10
  realizations corresponding to each $\omega$.}
\label{fig1_first}
\end{figure}

Let's first define the file density at a point $x$ of the line at
time $t$ as

\be
\rho(x,t)=\frac{n(x,t)}{dx}
\label{density}
\ee

\noi where  $n(x,t)$ refers to the number of particles in the bin
$\left[x-\frac{dx}{2},x+\frac{dx}{2}\right]$ at time $t$. It is
  straightforward to note that the quantity $\rho(x,t)$ is a \emph{local}
  property of the file, independent of the relabeling of the particles due to
  collisions. A direct consequence of this is that  the time evolution
  of the file profile density 
  can be  described by the Diffusion noise equation for a 1
  dimensional 
  gas of N non-interacting Brownian particles \cite{vanVliet}

\be
\frac{\partial}{\partial t}\rho(x,t)=-\frac{\partial}{\partial
  x}J(x,t),
\label{DN}
\ee

\noi having a recourse to the definition of a stochastic flux
$J(x,t)=-D\frac{\partial}{\partial  x}\rho(x,t)+\eta(x,t)$.
The noise term $\eta(x,t)$ can be shown to  be Gaussian and satisfy
the following properties 
\cite{VKampen}:

\be
\begin{array}{l}
\eta(L,t)=\eta(0,t)\\
\langle \eta(x,t)\rangle=0,\\
\langle
\eta(x,t)\eta(x',t')\rangle=2D\delta(x-x')\delta(t-t')\langle\rho(x,t)\rangle
\label{noiseproperty},
\end{array}
\ee

\noi the first one of which refers to the conservation of the
particle's number (\emph{conserved noise}) along the segment $[0,L]$
with periodic boundary conditions. The remaining properties in
(\ref{noiseproperty})  are required to fulfill the equations
for the first two moments of $\rho(x,t)$

\be
\begin{array}{l}
\frac{\partial}{\partial t}\langle\rho(x,t)\rangle  =  D\frac{\partial^2}{\partial
  x^2}\langle\rho(x,t)\rangle\\
\begin{split}
\left[\frac{\partial}{\partial t}-D\frac{\partial^2}{\partial
  x_1^2}-D\frac{\partial^2}{\partial
  x_2^2}\right]\langle\rho(x_1,t)\rho(x_2,t)\rangle_C  =\\
2D\frac{\partial}{\partial x_1}\frac{\partial}{\partial x_2}\delta(x_1-x_2)\langle\rho(x,t)\rangle,
\end{split}
\end{array}
\label{densitymoments}
\ee

\noi where we made use of the short notation
$\langle\rho(x_1,t)\rho(x_2,t)\rangle_C=\langle\rho(x_1,t)\rho(x_2,t)\rangle-\langle\rho(x_1,t)\rangle
\langle\rho(x_2,t)\rangle$.  As it is apparent from
(\ref{noiseproperty}), the correlation function of the noise depends upon
the particular solution  of the diffusion equation in
(\ref{densitymoments}): in the following, as well as in the numerical
simulations performed, we consider the case of a uniformly distributed
file, namely $\langle\rho(x,0)\rangle\equiv\langle\rho(x,t)\rangle\equiv\rho$.

The connection between the dynamics of the density over $[0,L]$ and
the the motion of a tagged particle in the single file system  is
achieved in the following way. Given two particle trajectories, $x_1(t)$
and $x_2(t)$, the number of
particles between these has to remain constant in time because of  the
non-overlapping condition, this implies 

\be
\frac{d}{dt}\int_{x_1(t)}^{x_2(t)}\rho(x,t)\,dx=0.
\label{condensity}
\ee

\noi Performing the derivative  and making use of (\ref{DN}), the
previous relation then reads

\[
\begin{split}
v_2(t)\rho\left(x_2(t),t\right)-J\left(x_2(t),t\right)-v_1(t)\rho\left(x_1(t),t\right)\\
+J\left(x_1(t),t\right)=0;
\end{split}
\]

\noi now, both the terms  must  set to
zero irrespective of the particle labels: we can thus write down the
equation for the single file particle as

\be
v(t)\rho\left(x(t),t\right)=J\left(x(t),t\right).
\label{SFeq}
\ee

\noi We notice that, although  the  relation (\ref{SFeq}) just
introduced is exact, it
is highly   
nonlinear. We are thus compelled to fall back upon two approximations in
order to solve it.  The first approximation is to 
assume that the density surrounding the particle position is
essentially constant: 
$\rho(0,t)\equiv\langle\rho(0,t)\rangle\equiv\rho\,$.
The second approximation we put forward is to assume that the particles movements are correlated over a range equal to the displacement of the tagged particle, such that we can take the current $J(x(t),t)$ to be equal to the current at the particles initial position. Taking this to be at $x=0$ we thus have $J(x(t),t)=J(0,t)$.
Notice that similar assumptions have been
superimposed by Alexander and Pincus in a previous treatment of the
single file subdiffusive dynamics on a lattice \cite{Pincus}. 

\begin{figure}[t]
\includegraphics[width=6cm]{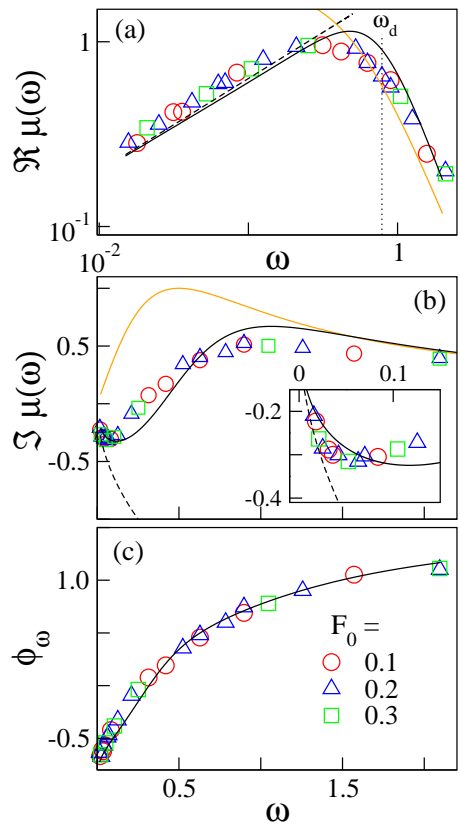}
\caption{(Color online). Real (a) imaginary  part (b) and phase (c) of
  $\mu{(\omega)}$  
  obtained through the relation (\ref{LinearResponse}), for different
  $F_0$ and $\omega$. Simulation parameters are the same as in 
  Fig.\ref{fig1_first}. The dashed lines represent the formula
  (\ref{DNmobility}) and are seen to fit quite well the asymptotic
  subdiffusive behavior of the mobility (see the inset of panel
  (b) where the low-frequency behavior of $\Im \mu (\omega)$ is blown
  up). The initial  
  diffusive behavior (orange lines) is responsible 
  for the high-frequency regime $\omega\gg\omega_d$ (dotted
  line in (a)). The GLE expression for the mobility
  (\ref{GLE_LT_mobility}) furnishes a very good description of both
  regimes: solid black lines. The non-perfect agreement between
  theory and data is due to the small value of the ratio
  $\tau_d/\tau_b=4$ (see text and TABLE I).}
\label{fig1}
\end{figure}

\noi With these approximations Eq.(\ref{SFeq}) gets the form

\be
v(t)\simeq \frac{J(0,t)}{\rho}.
\label{finalSFeq}
\ee

\noi Moreover defining the Fourier transform (and its inverse) in the
space and time domains as

\be
\begin{array}{l}
f(q,\omega)=\int_{-\infty}^{+\infty}dx dt\,
f(x,t)\,e^{-i(qx-\omega t)}\\ 
f(x,t)=\int_{-\infty}^{+\infty}\frac{dq
  d\omega}{\left(2\pi\right)^2}\,f(q,\omega)\,e^{i(qx-\omega t)},
\label{FT}
\end{array}
\ee

\noi the equation (\ref{finalSFeq}) can be casted as

\[
v(\omega)\simeq
\frac{1}{\rho}\int_{-\infty}^{+\infty}\frac{dq}{2\pi}J(q,\omega),
\]

\noi which, by means of (\ref{DN}) and of the definition of the
stochastic flux, reads

\be
v(\omega)\simeq
\frac{1}{\rho}\int_{-\infty}^{+\infty}\frac{dq}{2\pi}\,\frac{-i\omega}{q^2D-i\omega}\,\eta(q,\omega).
\label{FTSFeq}
\ee
 
Let us now  define the following quantity

\be
\mu(\omega)= \frac{1}{\rho\gamma}\int_{-\infty}^{+\infty}\frac{dq}{2\pi}\,\frac{-i\omega}{q^2D-i\omega};
\label{defDNmobility}
\ee

\noi using (\ref{FTSFeq}) and the noise properties  in
(\ref{noiseproperty}), it is readily verified that the following
equality holds:

\be
\langle
v(\omega)v(\omega')\rangle=2k_BT\,\Re\left[\mu(\omega)\right]\,2\pi\delta(\omega+\omega').
\label{spettroVCF}
\ee

\noi Furthermore a direct calculation of (\ref{defDNmobility}) gives

\be
\mu(\omega)=
\sqrt{\frac{2}{k_BT\,\gamma}}\,\frac{\sqrt{\omega}}{4\rho}\,[1-i] 
\label{DNmobility}
\ee

\noi which, substituted in (\ref{spettroVCF}) and consistent with the
Wiener-Khintchine theorem, yields  the asymptotic
form of the velocity autocorrelation function (VAF) of a single file
particle \cite{Marchesoni} 

\be
\langle
v(t)v(0)\rangle=-\sqrt{\frac{k_BT}{\gamma\,\pi}}\,\frac{1}{4\rho}\,\frac{1}{t^{3/2}}.
\label{VAF}
\ee

\noi  In passing from eq.(\ref{spettroVCF}) to (\ref{VAF}) we
implicitly adopted  the convention $\langle v(t)v(0)\rangle=\langle
v(-t)v(0)\rangle$, 
which uniquely determines the mobility
$\mu(\omega)$ to be the Fourier-Laplace transform of the VAF, i.e.

\be
\mu(\omega)=\frac{\int_0^{+\infty}\langle v(t)v(0)\rangle\,e^{i\omega t}}{k_BT}.
\label{1KT}
\ee

The relation (\ref{1KT}) is known as the first Fluctuation-Dissipation theorem
or Green-Kubo relation \cite{Kubo}. However for linear systems the
connection between  
correlation functions and mobility hinges on the domain of transport
processes \cite{Zwanzig-Transport}: within this framework the
equations (\ref{FTSFeq}) and 
(\ref{defDNmobility}) 
on one hand, and their statistical counterpart (\ref{1KT}) on the other,
encouraged us to explore the validity of a linear response relation of
the type

\be
\langle v(t)\rangle=\Re\left[\tilde{\mu}(\omega)\,F_0 e^{i\omega t}\right].
\label{LinearResponse}
\ee

\noi Indeed, applying a periodic external force $F(t)=F_0\cos\omega t$
to a tagged  particle and leaving  the 
remaining surrounding ones unaffected (see Fig.\ref{fig1_first}(a)),
the response  (\ref{LinearResponse}) has 
been directly 
measured by meaning of extensive numerical simulations.  The 
results are shown  in  Fig.\ref{fig1_first}(b) and Fig.\ref{fig1}, where
the real,  
the imaginary part and the relative phase $\phi_\omega$ of the
quantity $\tilde{\mu}(\omega)$ are 
displayed. At first we note as 
the average velocity depends linearly upon the force amplitude $F_0$
according to the equation (\ref{LinearResponse}). Most important, 
the low frequency behavior of  $\tilde{\mu}(\omega)$ fully agrees with the
analytical formula for the mobility $\mu(\omega)$ given in
(\ref{DNmobility}) (dashed line). We can interpret this result
recalling that a 
particle undergoes a normal diffusive behavior up to a time scale
$\tau_d=\frac{1}{D\rho^2}$, which is the time needed by a couple of
particle to collide with 
one of the neighboring diffusants \cite{Marchesoni}; correspondingly
we can assume that the tagged driven particle will \emph{feel}  the
presence of the surrounding ones over frequencies smaller than the
threshold $\omega_d=\frac{2\pi}{\tau_d}$ (dotted line). Conversely, for 
$\omega\gg\omega_d$, $\tilde{\mu}(\omega)$  coincide with the
mobility of a free Brownian walker: $1/(\gamma-i\omega)$. The
numerical results  in 
Fig.\ref{fig1} can thus be 
summarized by writing, besides the relation (\ref{LinearResponse}),
the mobility  $\tilde{\mu}(\omega)$  as

\be
\tilde{\mu}(\omega)\simeq\left\{
\begin{array}{ccc}
\frac{\gamma+i\omega}{\gamma^2+\omega^2} & & \omega\gg\omega_d\\
\sqrt{\frac{2}{k_BT\,\gamma}}\,\frac{\sqrt{\omega}}{4\rho}\,[1-i] & & \omega\ll\omega_d.
\label{LRmobility}
\end{array} 
\right.
\ee

The numerical evidence of the effectiveness of a linear
response relation  allows us to rewrite the eq.(\ref{FTSFeq}) as

\be
v(\omega)=\tilde{\mu}(\omega)\tilde{\xi}(\omega)
\label{FTSFeq_LRT}
\ee

\noi where the introduced noise  satisfies $\langle
\tilde{\xi}(t)\rangle=0$. On the other hand, its spectrum $S_\xi(\omega)$ exhibits two
different  
regimes according to (\ref{LRmobility}): 

\be
S_\xi(\omega)=\int_{-\infty}^{+\infty}\langle\xi(t)\xi(0)\rangle
e^{i\omega t}\simeq\left\{
\begin{array}{ccc}
2k_BT\gamma & & \omega\gg\omega_d\\
\frac{\left(2k_BT\right)^{3/2}\sqrt{\gamma}\rho}{\sqrt{\omega}}  & & \omega\ll\omega_d.
\label{Spectrumnoise}
\end{array} 
\right.
\ee

\noi The  file particles surrounding the tagged one thus act as an
additional bath responsible for the onset of
subdiffusional behavior. The nature of this long-ranged 
correlations
in the noise source can be easily understood in terms of the collisional
interaction between the file components. Note in fact that the
expression (\ref{Spectrumnoise}) leads to a slowly decaying positive
correlated 
noise  $\langle\tilde{\xi}(t)\tilde{\xi}(0)\rangle\propto\frac{1}{\sqrt{t}}$:
this non-trivial finding is the signature of constrained geometry
systems. Indeed, although  the  collisions tend to tie back the
particle motion, leading to the negative velocity correlations
(\ref{VAF}), the noise 
provides the way to maintain $\langle v(t)\rangle=0$ on time scale of
the order $1/\gamma$. Another way to 
say this, is that in a collision a particle exchanges velocity
\emph{and} noise. Anyway, this is a manifestation of the
Fluctuation-Dissipation theorem.
\noi Furthermore, the
non-Markovian power-law nature  
of the noise spectrum  characterizes the asymptotic Fractional
Brownian Motion (FBM) of the  tagged particle \cite{Mandelbroot}.

We end this section stressing that, using the formalism so far
developed, we can calculate the other asymptotic statistical
properties of the system. In fact it is straightforward to write down an
expression  for the 
particle's position similarly to  what we did in (\ref{FTSFeq}) for
its velocity:

\be
x(\omega)\simeq
\frac{1}{\rho}\int_{-\infty}^{+\infty}\frac{dq}{2\pi}\,\frac{\eta(q,\omega)}{q^2D-i\omega}.
\label{FTSFXeq}
\ee

\noi For instance, making use of (\ref{FTSFXeq}) and (\ref{FTSFeq}) we
get

\be
\langle
x(\omega)v(\omega')\rangle=\sqrt{\frac{k_BT}{2\gamma}}\frac{1}{\rho}\frac{i}{\sqrt{\omega}}\,2\pi\delta\left(\omega+\omega'\right).
\label{FTXVcorrfunc}
\ee

\section{Generalized Langevin description}
\label{sec:Generalized_Langevin}

In this section we will collect the results outlined in the
previous one and will put them in a consistent compact formulation. 
We emphasize three fundamental properties that such a representation must
incorporate

\emph{i)} the Linear Response relation (LR) must hold: (\ref{LinearResponse});

\emph{ii)} the Fluctuation Dissipation theorem (FDT) is also valid: (\ref{Spectrumnoise});

\emph{iii)} all the statistical properties exhibit two different
behaviors, Brownian or subdiffusive motion, depending on whether time is smaller or larger than $\tau_d$. 
We will in the following assume that the time scale $\tau_b=1/\gamma$ is smaller than $\tau_d$ such that the particles are moving diffusively (in contrast with ballistically) before colliding with each other.

 Our aim is thus to write down  an effective equation for the
microscopic dynamics of a tagged particle in single file systems,
according to $i),\,ii),\,iii)$.
Such an equation turns out to be

\be
\begin{array}{l}
\dot{x}(t)  =  v(t)\\
\dot{v}(t)  =  -\left[\gamma +2\,n_d^2\,\gamma_{1/2}\, _0D_t^{-1/2}\right]v(t)+\tilde{\xi}(t).
\label{GLE}
\end{array}
\ee

\noi Several new symbols have been introduced in the previous
expression  deserving an explanation. Firstly the quantity
$n_d=\frac{\gamma}{\rho\sqrt{k_BT}}$ plays the same role as $\tau_d$
in the collisional representation of the particle's motion
\cite{Marchesoni}: it accounts for the number of collisions  a
particle suffers before attaining subdiffusive behavior.  
Secondly the quantity $\gamma_{1/2}$, which  has the dimension of
$\left[1/t^{3/2}\right]$,  is the \emph{generalized damping
  coefficient} and
is equal to $\left(\frac{1}{\tau_d}\right)^{3/2}$ since $\tau_d$ is
the unique relevant  
time scale of the system. The third symbol in (\ref{GLE}) is the
Riemann-Liouville 
fractional operator \cite{Samko}:

\be
_0D_t^{-1/2}f(t)=\frac{1}{\Gamma\left(\frac{1}{2}\right)}\int_0^t\frac{f(t')}{\left|t-t'\right|^{\frac{1}{2}}}\,dt'. 
\label{RL}
\ee

\noi Note that if we use the definition of the Caputo fractional
derivative ~\cite{Caputo} 

\[
\frac{d^{1/2}f(t)}{dt^{1/2}}=\frac{1}{\Gamma\left(\frac{1}{2}\right)}\int_0^t\frac{df(t')/dt'}{\left|t-t'\right|^{\frac{1}{2}}}\,dt',
\]

\noi then (\ref{GLE}) takes the form

\be
\frac{d^2x(t)}{dt^2} + \gamma \frac{dx(t)}{dt}+2\,n_d^2\,\gamma_{1/2}\,\frac{d^{1/2}x(t)}{dt^{1/2}} =\tilde{\xi}(t).
\label{C_GLE}
\ee

\begin{figure}[t]
\includegraphics[width=5cm]{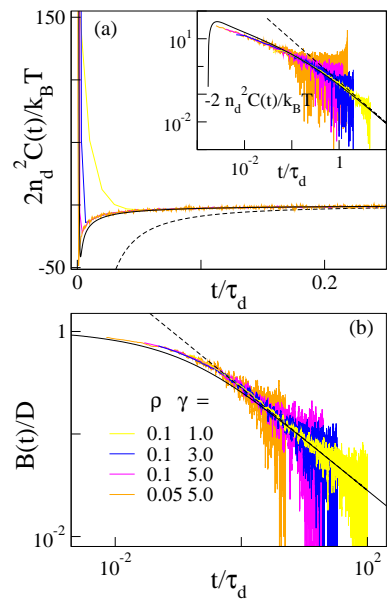}
\caption{(Color online) Second moments of observables: Rescaled VAF
  $\langle v(t)v(0)\rangle$ (a) and position-velocity correlation
  function $\langle\left[ x(t)-x(0)\right]v(t)\rangle$ (b)  
  \emph{vs} $t/\tau_d$ for several densities $\rho$ and damping
  $\gamma$. $k_BT=1.0$. Dashed lines are the asymptotic formulae
  (\ref{VAF}) and (\ref{FTXVcorrfunc}) respectively, whereas the solid ones
  represent the GLE predictions (\ref{GLE_1KT}) and
  (\ref{XV_corr_func}) for $\rho=0.05$ and $\gamma=5.0$. To improve our
  statistics we averaged over 3000 
  different realizations.}
\label{fig2}
\end{figure}

 The noise appearing in  (\ref{GLE}) satisfies the properties 
(\ref{Spectrumnoise}). Indeed we recall that the diffusive microscopic
time scale $\tau_b=\frac{1}{\gamma}$ can be expressed as
$\tau_b=\frac{\tau_d}{n_d^2}$ \cite{Marchesoni} so that the equation
(\ref{GLE}) can 
be re-casted as

\be
\dot{v}(t)  = -\int_0^t\tilde{\gamma}\left(t-t'\right)v(t')\,dt'+\tilde{\xi}(t),
\label{GLE_short}
\ee

\noi introducing the  generalized damping and defining the $\delta$
function to contribute half at the end point of an integral:

\be
\tilde{\gamma}(t)=2\,\gamma\left[\delta(t)+\frac{1}{\sqrt{\pi\,\tau_d\,t}}\right].
\label{generalizeddamping}
\ee

\noi Thence such a definition allows us to express the properties in
(\ref{Spectrumnoise}) through the compact notation of the
\emph{Generalized Fluctuation-Dissipation theorem} \cite{Kubo}

\be
\langle\tilde{\xi}(t)\tilde{\xi}(t')\rangle=k_BT\,\tilde{\gamma}(\left|t-t'\right|)  
\label{GFDT}
\ee

The solution of the equation (\ref{GLE}) can be easily achieved by
means of the Laplace transform:

\be
\begin{array}{l}
x(s)  =  \frac{x(0)}{s} + v(0)\tilde{\psi}(s) + \tilde{\xi}(s)\tilde{\psi}(s)\\
v(s)  =  v(0)\tilde{\mu}(s) + \tilde{\xi}(s)\tilde{\mu}(s),
\label{GLE_LTsol}
\end{array}
\ee

\noi where the mobility $\tilde{\mu}(t)$ in the $s-$domain is given by 

\be
\tilde{\mu}(s)=\frac{1}{s+\tilde{\gamma}(s)}=\frac{1}{s+\gamma+2\,n_d^2\,\gamma_{1/2}\,s^{-1/2}}
\label{LTmobility}
\ee

\noi and

\be
\tilde{\psi}(s)=\frac{\tilde{\mu}(s)}{s}=\frac{1}{s^2+\gamma s+2\,n_d^2\,\gamma_{1/2}\,s^{1/2}}.
\label{LTpsi}
\ee

\noi It is immediate to verify that the expression in (\ref{LTmobility})
matches the predicted two regimes in (\ref{LRmobility}). In the time
domain the solution of the Generalized Langevin Equation in
(\ref{GLE}) takes the simple form

\be
\begin{array}{l}
x(t)  =  x(0) + v(0)\tilde{\psi}(t) + \int_0^t\tilde{\xi}(t')\tilde{\psi}(t-t')\,dt'\\
v(t)  =  v(0)\tilde{\mu}(t) + \int_0^t\tilde{\xi}(t')\tilde{\mu}(t-t')\,dt',
\label{GLE_sol}
\end{array}
\ee

\noi from which it is apparent that the joint probability
distribution for $x$ and $v$ is a Gaussian in this formulation. Position and velocity are
in fact linear functionals of $\tilde{\xi}(t)$ which is a (non-Markovian)
Gaussian random 
process. Such a property corroborates  all the previous analytical
derivations for the probability distribution of a tagged particle moving subdiffusively in
a single file system (see for example Ref.\cite{SF_basic} and
references therein) and it is clearly at odds with a Fractional
Fokker-Planck  description of the corresponding subdiffusive process,
which leads to a stretched Gaussian solution
\cite{Klafter}. 

\noi Furthermore we
point out that the GLE (\ref{GLE}) correctly describes
 the time behavior of  all the observable moments of position and
 velocity, not only in the long time asymptotic regime but also at the initial
 diffusive stage. To see this it is sufficient to start from the general
 solutions (\ref{GLE_sol}) and put in some physical assumptions
 \cite{Mazo}.
For instance the first moments of velocity and position will read

\be
\begin{array}{l}
\langle x(t)\rangle  =  \langle x(0)\rangle + \langle v(0)\rangle\tilde{\psi}(t)\\
\langle v(t) \rangle =  \langle v(0)\rangle \tilde{\mu}(t).
\label{GLE_1moments}
\end{array}
\ee

\noi The second moments, however, are more interesting quantities: for
the VAF, assuming $\langle v^2(0)\rangle=k_B T$,  it is
straightforward to prove 

\be
C(t)\equiv\langle v(t) v(0)\rangle=k_BT\,\tilde{\mu}(t)
\label{GLE_1KT} 
\ee

\noi which is the \emph{Generalized First Kubo theorem}. 
The numerical evidence of this is given in Fig.{\ref{fig2}}(a), where
several rescaled curves are plotted against the analytic function in
(\ref{GLE_1KT})  after numerical inversion of the mobility in
(\ref{LTmobility}).

\begin{figure}[t]
\includegraphics[width=7cm]{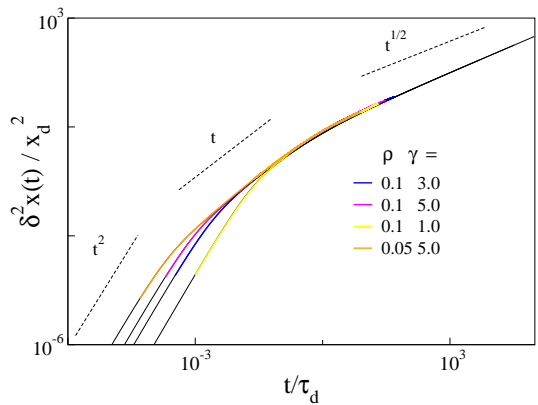}
\caption{(Color online) Mean square displacement $\delta
  x^2(t)=\Sigma_{i=1}^N\left[x_i(t)-x_i(0)\right]^2/N$  for
  different values of $\rho$ and $\gamma$ with 
  $k_BT=1.0$ plotted versus $t/\tau_d$. Data have been rescaled by
  $x_d^2=2/\left(\rho^2\sqrt{\pi}\right)$ on y axis \cite{Marchesoni}. The
  black solid lines represent the theoretical predictions as given by the
  GLE formula (\ref{MSD}): they provide an 
  excellent description of  the
  three diffusive regimes. Data have been averaged over 30 different
  realizations for \emph{N} of the order of $10^4$.
  }
\label{fig3}
\end{figure}

The excellent agreement between the analytical description yielded by
(\ref{GLE}) 
and the numerical data is even more apparent looking at the mean
square displacement of the tagged particle (Fig.\ref{fig3}). Indeed
from (\ref{GLE_sol}) 
one obtains 

\be
\delta x^2(t)\equiv\langle \left[x(t)-x(0)\right]^2\rangle=2\,\langle v^2(0)\rangle \int_0^t
\tilde{\psi}(t')\,dt',
\label{MSD}
\ee

\noi  where it is possible to recognize the theorem stated in
Ref.\cite{Marchesoni}   provided that 

\be
\frac{d}{dt}\tilde{\psi}(t)=\tilde{\mu}(t).
\label{psi_eq}
\ee

\noi The exact expression given in
(\ref{MSD}) quantitatively reproduces the \emph{three} stages of the
$\delta x^2(t)$ curves in Fig.\ref{fig3} (ballistic,
diffusive, subdiffusive) as well as the  characteristic
time scales ($\tau_b$ and $\tau_d$) on which the crossovers between
them take place
\begin{equation}
\delta x^2(t)=\left\{
\begin{array}{ccc}
k_BT\,t^2 & & t\ll\tau_b\\
2D\,t & & \tau_b\ll t\ll\tau_d\\
2\sqrt{\frac{ D}{\pi\rho^2}}\sqrt{t} & & t\gg\tau_d.
\label{SFstocregimes}
\end{array}
\right.,
\end{equation}
An analytical inversion of the formula in
(\ref{LTpsi}) including the crossover to the subdiffusive regime can be achieved by neglecting the inertial (ballistic)
term leading to
\be
\langle
\left[x(t)-x(0)\right]^2\rangle\simeq\frac{1}{\rho^2}\left[\sqrt{\frac{4t}{\pi\tau_d}}+\frac{e^{\frac{4t}{\tau_d}}}{2}\,\mathrm{erfc}\left(\sqrt{\frac{4t}{\tau_d}}\right)-\frac{1}{2}\right]
\label{MSD_appr}
\ee

\noi
In general we believe the
GLE (\ref{GLE}) provides a very good description of the tagged particle's
stochastic motion only when  the condition  $\tau_b\ll\tau_d$ is
fulfilled (see TABLE I). In other words the particle must attain a truly  
diffusive regime before getting a collision with one of its nearest
neighbors. One can see that the proposed GLE cannot work well in the opposite case where $\tau_b>\tau_d$ since all interactions with neighboring particles vanish according to Eq. (\ref{GLE}) as $\gamma\to 0$. This cannot be true, since the particle will still collide and exchange momentum with its neighbors even though the rest of the friction with the surroundings vanish. We have not found a GLE that works well in the case $\tau_b>\tau_d$.

\begin{table}[t]
\begin{tabular}{|c|c|cc|c|}
	\hline
$\rho$ &  $\gamma$    &   $\tau_b$  & $\tau_d$   & $\tau_d/\tau_b$  \\
	\hline
0.25     & 0.5        & 2.0         & 8.0        &  4             \\
0.1      & 1.0        & 1.0         & 100        &  100           \\
0.1      & 3.0        & $\sim$0.334 & 300        &  900           \\
0.1      & 5.0        & 0.2         & 500        &  2500          \\
0.05     & 5.0        & 0.2         & 2000       &  10000         \\
	\hline
\end{tabular}
\caption{Values of $\tau_{\rm b}$, $\tau_{\rm d}$ and their ratio for the values of $\rho$ and $\gamma$ used in Figs. \ref{fig1}, \ref{fig2}, \ref{fig3}, and \ref{fig5}.}
\label{table}
\end{table}

\noi The close comparison between theory and numerics when $\tau_b \ll \tau_d$ is also
 displayed by the position-velocity correlation function $\langle
 x(t)v(t)\rangle$ for which  the following expression holds

\be
B(t)\equiv\langle\left[x(t)-x(0)\right]v(t)\rangle=\langle v^2(0)\rangle\,
\tilde{\psi}(t)
\label{XV_corr_func}
\ee

\noi In Fig.\ref{fig2}(b) we compare the numerical data with both
(\ref{XV_corr_func}) and (\ref{FTXVcorrfunc}), which is
expected to work well only in the asymptotic regime.

\begin{figure}
\includegraphics[width=6cm]{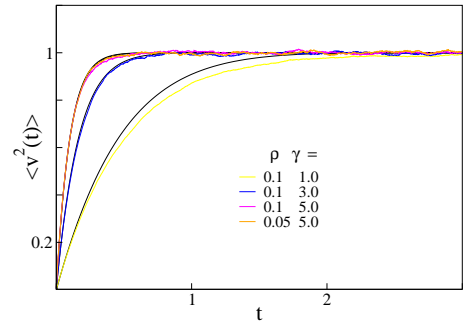}
\caption{(Color online). Second moment of the velocity $\langle
  v^2(t)\rangle$ \emph{vs} $t$, 
  displayed for several densities $\rho$    
  and damping $\gamma$. $k_BT=1.0$. The initial  tagged particle velocity $
  v(0)$ has been set to 0. The expression (\ref{V2t}) (solid black
  lines) shows the typical  Brownian exponential relaxation to $k_BT$:
  for all the displayed curves $\tau_b/\tau_d\ll 1$. To improve our
  statistics we averaged the numerical curves over 3000 realizations.}
\label{fig5}
\end{figure}

\noindent The relaxation of the second moment of the velocity $\langle
v^2(t)\rangle$ to the asymptotic value $k_BT$ is given by

\begin{equation}
\langle v^2(t)\rangle=\langle v^2(0)\rangle\,\mu^2(t)+k_BT\,\left[1-\mu^2(t)\right],
\label{V2t}
\end{equation}

\noindent and deserves particular attention. In Ref.~\cite{Lutz} it has
been shown that the Fractional Brownian process ~\cite{Mandelbroot}
generated by a Fractional Langevin equation and the
stochastic process corresponding to the relative  Fractional Kramer
equation   ~\cite{Barkay} are \emph{on average} the same,
except for 
the second moment of the velocity. We point out that such a
discrepancy is a common problem in non-equilibrium statistical
mechanics, whenever one is concerned to pass from a Generalized
Langevin dynamical description of a non-Markovian process ~\cite{Mori}  to the
corresponding Fokker-Planck equation for the probability
distribution ~\cite{Zwanzig} (see for instance the discussion in
Ref.~\cite{Chang} on the inconsistency of a retarded Fokker-Planck
equation of the Rubin model ~\cite{Rubin}).
In the GLE (\ref{GLE}) the  $\langle v^2(t)\rangle$
relaxation process
is dictated by the Brownian dynamics (see Fig.~\ref{fig5}) as long
as $\tau_b\ll\tau_d$. The Fractional Kramer equation
corresponding to 
(\ref{GLE})
would read

\begin{eqnarray}
&&\frac{\partial P(x,v,t)}{\partial t} + v\,\frac{\partial
  P(x,v,t)}{\partial
  x}\nonumber\\
&&\quad\quad =\left[\gamma+\gamma_{1/2}\,_{0}D_{t}^{-1/2}\right]\hat{L}_{FP} P(x,v,t),
\label{GFP}
\end{eqnarray}

\noindent where $\hat{L}_{FP}=\frac{\partial}{\partial
  v}v+k_BT\frac{\partial^2}{\partial v^2}$. In the  $\tau_b\ll\tau_d$
limit the $\langle v^2(t)\rangle_{FKE}$ expression yielded by (\ref{GFP}) is
the one provided by the usual Fokker-Planck, i.e. $\langle
v^2(t)\rangle_{FKE}\simeq k_BT\,e^{-2\gamma t}$ once $
v(0)=0$, in agreement with 
the exponential saturation shown in Fig.~\ref{fig5}. This could lead
to believe that the GLE description
(\ref{GLE}) and the Kramer Eq.(\ref{GFP}) are \emph{de
  facto} equivalent  in the range
$\tau_b\ll\tau_d$. Nevertheless 
we expect that the $P(x,t)$ solution  of (\ref{GFP}) is still a stretched
exponential instead of a Gaussian: this fundamental difference casts
some general  doubt on the
possibility to determine a Fokker-Planck equation for systems whose
microscopic dynamics is represented by a FLE.

The last part of this section is devoted to the property $i)$, namely
the validity of LR. In the presence of an external force acting from time equal zero and onwards, (\ref{GLE})-(\ref{GLE_short})  takes
the form

\be
\begin{array}{l}
\dot{x}(t)  =  v(t)\\
\dot{v}(t)  =  -\int_0^t\tilde{\gamma}\left(t-t'\right)v(t')\,dt'+F(t)+\tilde{\xi}(t).
\label{GLE_tilted}
\end{array}
\ee
\noi The LR  (\ref{LinearResponse}) in $s-$space can thus expressed by
using (\ref{LTmobility}) as

\be
\langle v(s)\rangle=\tilde{\mu}(s)F(s),
\label{GLE_LRT}
\ee

\noi so that , thanks to the expression (\ref{LTmobility}), the real and
imaginary part 
of the mobility $\tilde{\mu}(t)$ read

\be
\begin{array}{l}
\Re\left[\tilde{\mu}(\omega)\right]=\frac{\gamma+\sqrt{2}\,n_d^2\,\gamma_{1/2}\omega^{-1/2}}{\left(\gamma+\sqrt{2}\,n_d^2\,\gamma_{1/2}\omega^{-1/2}\right)^2+\left(\sqrt{2}n_d^2\,\gamma_{1/2}\omega^{-1/2}-\omega\right)^2}\\
\Im\left[\tilde{\mu}(\omega)\right]=\frac{\omega-\sqrt{2}n_d^2\,\gamma_{1/2}\omega^{-1/2}}{\left(\gamma+\sqrt{2}\,n_d^2\,\gamma_{1/2}\omega^{-1/2}\right)^2+\left(\sqrt{2}n_d^2\,\gamma_{1/2}\omega^{-1/2}-\omega\right)^2}
\label{GLE_LT_mobility}
\end{array}
\ee

\noi Remarkably, in Fig. \ref{fig1}  both  functions 
(\ref{GLE_LT_mobility}) (solid black lines) are shown to fit quite well the outcome of our
numerics.

\noi Finally in the case of a constant external force, i.e. $F(t)=F$ for $t\ge 0$ in
(\ref{GLE_tilted}), the LR provides that the drift
$\langle x(t)\rangle_F$ satisfies the
\emph{Generalized Einstein relation} 

\be
\langle x(t)\rangle_F=\langle x(0)\rangle+F\,\frac{\delta x^2(t)}{2\,k_BT}
\label{GER}
\ee

\noi that has been shown to be
valid on file systems moving on a ring \cite{Ferrari} and on an
infinite one-dimensional lattice
\cite{Oshanin}.

\section{Conclusion}
\label{sec:Conclusion}

In this paper we introduced the effective equation ruling  the
microscopic stochastic motion of a SF tagged particle. Starting from a
diffusion-like formalism for the file density dynamics, we were led to
several properties for which the GLE  (\ref{GLE})-(\ref{GLE_short}) can be
regarded as a representation.  Indeed the GLE formalism provides
an elegant representation of Generalized Fluctuation-Dissipation 
theorems (\ref{GFDT})-(\ref{GLE_1KT}), Linear
Response (\ref{GLE_LRT}) and  Generalized Einstein relation
(\ref{GER}), all remarkable properties satisfied by the particle,
both in its diffusive and in subdiffusive phase.

\noi Nevertheless we want to stress that the GLE is just a
description, though very good,  of the stochastic motion of the SF
particle and, in this perspective, is valid within certain
limits. However, along the same line, within certain approximations
the Langevin Equation provides an 
excellent  effective equation for the motion of a Brownian particle
 immersed in a thermal bath, i.e. it describes well the ballistic
 and diffusive regimes of the 
particle, but one would not expect it to be exact in the region of crossover
between the two regimes. In SF systems the tagged particle
is subjected to two types of  thermal baths, as is apparent from (\ref{GLE}):
the first is mimicked by the usual Markovian uncorrelated noise, whereas the
second, physically embodied by the surrounding file's particles,
is achieved by the introduction of 
an additional non-Markovian term, responsible for the strong memory
effects. Surprisingly this ``sum of thermal baths'' turns out to be
well described (in the same
manner as for the Brownian particle/LE) by simply
the  algebraic sum of two independent terms in the stochastic
equation of motion. Alternatively one could view 
(\ref{GLE}) as an usual  LE (\ref{LE}) where the
relabeling-collisional symmetry accounts for the fractional term.

\subsection*{Acknowledgments}

 We are indebted with Prof. Mehran Kardar for having deeply influenced
 this work. We also like to thank Prof. Fabio Marchesoni and  Tobias
 Ambj{\"o}rnsson for several useful
 discussions. A.T. is grateful to acknowledge  the ``Angelo Della
 Riccia'' foundation for having supported this work. M.A.L.
thanks the Danish National Research Foundation for support via a grant to
MEMPHYS - Center for Biomembrane Physics.


\end{document}